# Dual reciprocity BEM and dynamic programming filter for inverse elastodynamic problems

Masataka Tanaka[*] and Wen Chen[**]

Department of Mechanical Systems Engineering, Shinshu University, Wakasato 4-17-1, Nagano 380-8553, Japan (E-mail: [*]dtanaka@gipwc.shinshu-u.ac.jp; [**]chenw@homer.shinshu-u.ac.jp).

This paper presents the first coupling application of the dual reciprocity BEM (DRBEM) and dynamic programming filter to inverse elastodynamic problem. The DRBEM is the only BEM method, which does not require domain discretization for general linear and nonlinear dynamic problems. Since the size of numerical discretization system has a great effect on the computing effort of recursive or iterative calculations of inverse analysis, the intrinsic boundary-only merit of the DRBEM causes a considerable computational saving. On the other hand, the strengths of the dynamic programming filter lie in its mathematical simplicity, easy to program and great flexibility in the type, number and locations of measurements and unknown inputs. The combination of these two techniques is therefore very attractive for the solution of practical inverse problems. In this study, the spatial and temporal partial derivatives of the governing equation are respectively discretized first by the DRBEM and the precise integration method, and then, by using dynamic programming with regularization, dynamic load is estimated based on noisy measurements of velocity and displacement at very few locations. Numerical experiments involved with the periodic load are conducted to demonstrate the applicability, efficiency and simplicity of this strategy. The affect of noise level, regularization parameter, and measurement types on the estimation is also investigated.

## 1. Introduction

In recent years inverse elastodynamic problems have received increasing attention due to a broad range of engineering necessity. In general, the solution of the inverse dynamic problem is a much more difficult task than the direct problem due to some degree of noise in the measurement data. In other words, the inverse solution is extremely sensitive to measurement errors, namely, the ill-posed nature of the inverse problem [1,2]. In compared with direct problems, research of inverse dynamic problems is much less reported in literature, especially for inverse elastodynamic problem [3]. There are several methods available now to stabilize and estimate the inverse solutions of dynamic problem [2]. Among them, the dynamic programming filter with regularization, introduced recently by Trujillo and Busby [3,4], is a very competitive technique. The strengths of this approach lies in the mathematical simplicity, easy to program and its great flexibility in the type, number and location of measurements and unknown excitation sources.

On the other hand, an appropriate numerical method is also required in inverse analysis to transfer the continuous models of various practical problems into discretization system. In recent years, the BEM has become increasingly popular in the numerical discretization of dynamic partial differential systems occurring in many branches of science and engineering. Transformation of the domain integrals has been a central task in the BEM solution of such problems to preserve its boundary-only merit. There are several different approaches available now for this purpose. However, as was pointed out in [5,6], the dual reciprocity BEM (DRBEM) stands out the method of choice in engineering computations due to its ease of implementation, intrinsic boundary-only merit for general problems, meshless grids and strong flexibility of applying fundamental solutions. Much research has been reported in literature to apply the DRBEM to a variety of direct dynamic problems. In contrast, there has been only very limited amount of research carried out in the DRBEM analysis of inverse dynamic problems. It is also worth stressing that since the inverse analysis usually require recursive or iterative computation many times, the dimension of numerical analogous equations has an especially huge affect on the computing time and storage requirements. The boundary element method enjoys a far more saving in computer resources for inverse analysis in comparison with the domain-type method such as the standard FEM and FDM because the method produces a relatively much less size of numerical modeling for continuous system [7]. However, it is noted that the dynamic programming filter so far is only applied to analyze some structural dynamic problems in the combination with the FEM [3,8]. It was also claimed in [5,6] that the DRBEM is the only BEM method, which does not require domain discretization for general linear and nonlinear problems, although the interior collocation points may be used to improve solution in some cases. The relatively small dimension of the DRBEM discretization equations is especially advantageous for the dynamic programming filter, since its computational effort increases quickly as the size of system equation increases. A combined use of the DRBEM and dynamic programming filter will be very attractive in terms of computational efficiency compared with the FEM and other BEM techniques.

The purpose of this study is concerned with the estimation of the input force magnitude of elastodynamic problems by a combined use of the DRBEM and dynamic programming filter. As for the approximate method of time derivative, Zhong and Williams [9] recently presented a so-called precise integration method (PIM). The method is in fact equivalent to the exponential matrix approach used in [3]. The merit of the PIM over the latter is mathematically explicit and easy to use. In this study, we employ this technique to discretize temporal derivative. Numerical experiments are plates subjected to in-plane periodic load. Dynamic input load is estimated based on noisy measurements of velocity or displacement at very few locations. A computer program generating random number is employed to yield random measurement errors. The exact displacement and velocity responses versus time can be contaminated with various amount of noise to better simulate real measurements. The detailed solution procedure is next explained and some conclusions are finally drawn based on the present work. The main points of the interest are to investigate the affect of noise level, regularization parameter, and measurement types on the estimation. To the authors' best knowledge, this is the first attempt to use the DRBEM combined with dynamic programming filter to handle the inverse dynamic problem.

## 2. Numerical modeling of plate elastic wave

The cantilever plate subjected to in-plain dynamic load is often used as a benchmark problem in the BEM analysis [5] due to the availability of its analytical solution. In this study, we choose it as the numerical examples. The equation describing a wave propagating through an elastic medium is given by

$$\nabla^2 u(x,t) = \frac{1}{c^2}\frac{\partial^2 u(x,t)}{\partial t^2}, \quad x \in \Omega \qquad (1)$$

subjected to the initial conditions

$$u(x,0) = 0, \qquad (2a)$$
$$\dot{u}(x,0) = 0, \qquad (2b)$$

and the displacement and traction boundary conditions

$$u(x,t) = 0, \quad x_1 = 1, \qquad (3a)$$
$$T(x,t) = 0, \quad x_2 = 0, 1, \qquad (3b)$$
$$T(x,t) = P, \quad x_1 = 0, \qquad (3c)$$

where c denotes the wave velocity. $P$ is the external plain traction as shown in Fig. 1. Note that all variables are dimensionless in this study. The analytical solution of this problem can be found in [10]. In terms of BEM, Eq. (1) is weighted by the fundamental solution $u^*$ of Laplace operator, namely,

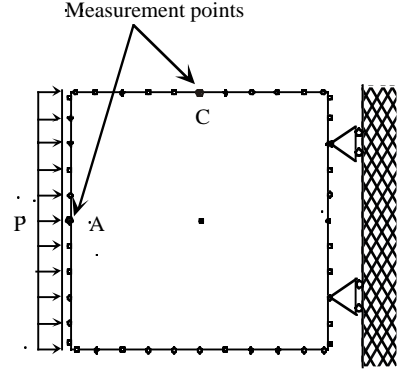

Fig. 1. DRBEM discretization and measurement points of a square plate

$$\int_\Omega \left(\frac{1}{c^2}\frac{\partial^2 u}{\partial t^2} - \nabla^2 u\right) u^* d\Omega = 0. \qquad (4)$$

By employing Green's second identity, we have

$$d_i u_i + \int_\Gamma (T^* u - u^* T) d\Gamma = -\int_\Omega \frac{1}{c^2}\frac{\partial^2 u}{\partial t^2} u^* d\Omega, \qquad (5)$$

where subscript $i$ denotes the source point, $T^* = \partial u^*/\partial n$, $n$ is the unit outward normal; and $d_i = \int_\Omega \delta(\zeta, x) d\Omega$. $\Gamma$ means the boundary of plate. The essence of the dual reciprocity BEM is to transform the domain integral on the right-hand side of Eq. (5) by a set of coordinate function $f^j(x)$

$$\ddot{u}(x,t) \approx \sum_{j=1}^{N+L} f^j(x) \ddot{\alpha}^j(t), \qquad (6)$$

where the superimposed dot represents the time derivative of the second order, $\alpha^j$ are unknown functions of time, and $N$ and $L$ are the numbers of the boundary and selected internal nodes, respectively. After some inferences, the DRBEM formulation is given by

$$d_i u_i + \int_\Gamma (T^* u - u^* T) d\Gamma = \\ \sum_{j=1}^{N+L} \left[d_i \psi_i^j + \int_\Gamma (T^* \psi^j - u^* \eta^j) d\Gamma\right]\frac{\ddot{\alpha}}{c^2}, \qquad (7)$$

where $\eta^j = \partial \psi^j/\partial n$, functions $\psi^j(x)$ are linked with the specified coordinate functions $f^j$ through

$$\nabla^2 \psi^j = f^j. \qquad (8)$$

The coordinate functions presented in [11] are applied in this study. The formulation (7) can be restated in matrix form as

$$M\ddot{u} + Hu - GT = 0, \quad (9)$$

where $M$ is the mass matrix, $H$ and $G$ denote the whole matrices of boundary element with kernels $T^*$ and $u^*$, respectively. All these coefficient matrices are dependent only on the geometric data of the problem. In this study, the linear element ($\Delta\Gamma$=0.1) is employed, and one internal point is placed in the interior domain as shown in Fig. 1.

Since displacement boundary conditions are involved in the cantilever plate, Eq. (9) is a differential algebraic system. By using an approach of matrix partition [5], the DRBEM formulation can be reduced to

$$m\ddot{u} + ku = f, \quad (10)$$

where $m$ and $k$ are respectively mass and stiffness matrices.

In what follows, we use the precise integration method to approximate time derivative. The above equation (10) can further be restated as the first-order system, namely,

$$\dot{v} = Hv + r, \quad (11)$$

in which

$$v = \begin{Bmatrix} u \\ \dot{u} \end{Bmatrix}, \quad H = \begin{bmatrix} 0 & m \\ -k & 0 \end{bmatrix}, \quad r = \begin{Bmatrix} 0 \\ I \end{Bmatrix} f. \quad (12)$$

$I$ is the unit matrix. The key step in the PIM is to accurately evaluate the exponential matrix

$$T = exp(H \bullet \tau) \quad (13)$$

by using

$$T(\tau) = \left[ exp(H\Delta\bar{t}) \right]^m, \quad (14)$$

where $\tau$ denotes the time step size, $\Delta\bar{t} = \tau/m$, and $m = 2^N$. $N$=20 is used to assure high accuracy of the matrix $T$. Therefore, $\Delta\bar{t}$ is extremely small time interval and usually much less than the highest modal period of dynamic systems. By using a Taylor expansion, we have

$$exp(H\Delta\bar{t}) \cong I + \left[ H\Delta\bar{t} + (H\Delta\bar{t})^2/2! + (H\Delta\bar{t})^3/3! + (H\Delta\bar{t})^4/4! \right] = I + T_{a,0}. \quad (15)$$

Substitution of Eq. (15) into Eq. (14) gives

$$T(t) = \left[ I + T_{a,0} \right]^{2^N}. \quad (16)$$

A recurrence procedure of computing $T$ is given by

$$T_{a,i} = 2T_{a,i-1} + T_{a,i-1} \times T_{a,i-1}. \quad (17)$$

Finally, we have

$$T = I + T_{a,N}. \quad (18)$$

The approximation in Eq. (18) is caused by the truncation of the Taylor expansion of Eq. (15). As was pointed out in [9], the truncation error is of the order $O(\Delta\bar{t}) = 10^{-30} O(\Delta\tau)$ under $N$=20, which is of the order of the round-off errors of ordinary computers. So it is claimed in [9] that the exponential matrix $T$ calculated by the PIM has the highest accuracy of a digital computer

After computing exponential matrix $T$, the general solution $v$ to Eq. (11) is given by

$$v_{j+1} = T\left[ v_j + H^{-1}\left( r_j + H^{-1}r_1 \right) \right] - H^{-1}\left[ r_{j+1} + H^{-1}r_1 \right], \quad (19)$$

where $r_1 = (r_{j+1} - r_j)/\tau$. Note that the outer forcing term has been assumed to vary linearly within time step $[t_j, t_{j+1}]$, i.e.,

$$r = r_j + r_1(t - t_j). \quad (20)$$

Eq. (19) can be restated as

$$v_{j+1} = Tv_j + Pr_j + D(r_{j+1} - r_j), \quad (21)$$

where $P = (T-I)H^{-1}$, $D = (P-\tau)H^{-1}$. The above equation (21) is the numerical discretization equation of the present elastodynamic problem.

## 3. Inverse problem solution

Trujillo and Busby [3] pointed out that the first-order regularization generally performs better than the zero-order one for the dynamic problems. The first-order regularization formulation for elastodynamic equation (21) can be given by

$$z_{j+1} = Rz_j + Gq_j. \quad (22)$$

where $q_j = r_{j+1} - r_j$,

$$R = \begin{bmatrix} T & P \\ 0 & I \end{bmatrix}, \quad G = \begin{bmatrix} D \\ I \end{bmatrix}, \quad z_j = \begin{Bmatrix} v_j \\ r_j \end{Bmatrix}. \quad (23)$$

$q$ is actually the first order derivative of the forcing term. The state vector $z$ now includes the forcing term. The essence of the dynamic programming filter [3,4] is to formulate an optimal control problem, namely,

$$E_N(z, q_i) = \sum_{i=1}^{N} \left( d_i - d_i^*, A(d_i - d_i^*) \right) + (q_i, Bq_i), \quad (24)$$

where $(x,y)$ denotes the inner product of two vectors, $N$ is the number of measurements, $d_i^*$ denote the measurement data, and $A$ is the weighting matrix and chosen as the identity matrix in this study. $d_i$ represent the state variable corresponding to the measurements. $B$ is the Tikhonov regularization parameter. The L-curve method and generalized cross validation are two approaches in use for selecting optimal regularization parameter. In this study, we apply the former. $d$ can be related to the state variable $z$ by

$$d_j = Qz_j. \tag{25}$$

By applying the least-squares criteria and dynamic programming principle to Eq. (24), we can get two sets of forward and backward recurrence formulas, respectively. The first step in computation is backward recurrence, namely,

$$D_n = \left(2B + 2G^T E_n G\right)^{-1}, \tag{26a}$$

$$F_n = 2G^T E_n, \tag{26b}$$

$$E_{n-1} = Q^T A Q + R^T \left(E_n - F_n^T D_n F_n / 2\right) R, \tag{26c}$$

$$s_{n-1} = -2Q^T A d_{n-1}^* + R^T \left(I - F_n^T D_n G\right) s_n. \tag{26d}$$

The above equations are solved with initial conditions starting at the end point $n=N$

$$E_N = QAQ^T, \tag{27a}$$

$$s_N = -2Q^T A d_N^*. \tag{27b}$$

Note that all vectors $D_n G^T s_n$ and matrices $D_n F_n R$ should be stored during the backward sweep. The forward solution is then calculated by the recurrence formulas

$$q_{n-1} = -D_n G^T s_n - D_n F_n R z \tag{28}$$

and Eq. (22) from $n=1$. One can find from the above recursive formulas that the dynamic programming filter is mathematically simple and easy-to-program. Also the method is not restricted to the numbers and locations of measurements and unknown input terms.

## 4. Numerical results and discussions

In this study, the plates subjected to the Heaviside impact and harmonic load are tested. The measured data are artificially produced by corrupting the exact displacement or velocity history at some sample points with different degree of noise, namely,

$$d_j^* = c_j + \varepsilon_j, \tag{29}$$

where $c_j$ is the exact response of velocity or displacement, $d_j^*$ represents the corresponding contaminated data and is considered as noisy measurements. $\varepsilon_j$ is the added noise generated by

$$\varepsilon_j = P \bullet A \bullet \left(\gamma_j - 0.5\right), \tag{30}$$

where $A$ is the peak velocity or displacement value of the forced vibration at measurement points, $\gamma_j$ is the normally-distributed random number over the interval [0,1] with zero mean generated by a computer routine [12]. $P$ is the noise percentage degree of the amplitude $A$. $PA$ actually denotes the given standard variance of random measurement noise.

The advantage of such numerical experiments is that the performances of the present methodology can easily be compared and evaluated with available analytical solutions. In the following, we will investigate the utility of the DRBEM combined with dynamic programming filter and observe whether the estimations are sensitive to the regularization parameter, noise level and the number and types of measurements. It is noted that dimensionless time step size $c\Delta t = 0.1$ is employed for all results discussed below.

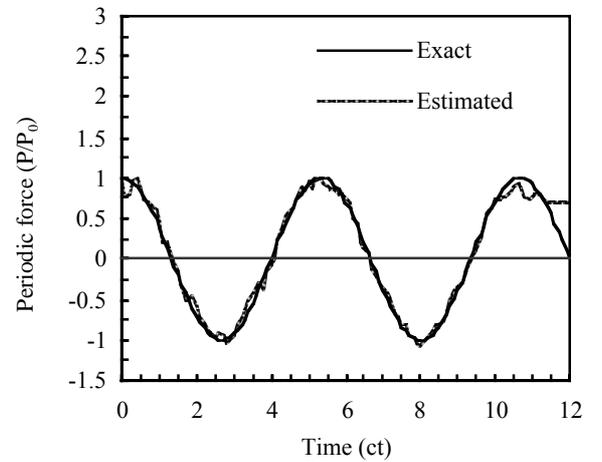

Fig. 2. Periodic load estimation based on velocity measurement of 5 percent noise level at one point $C$ ($B$=4.1)

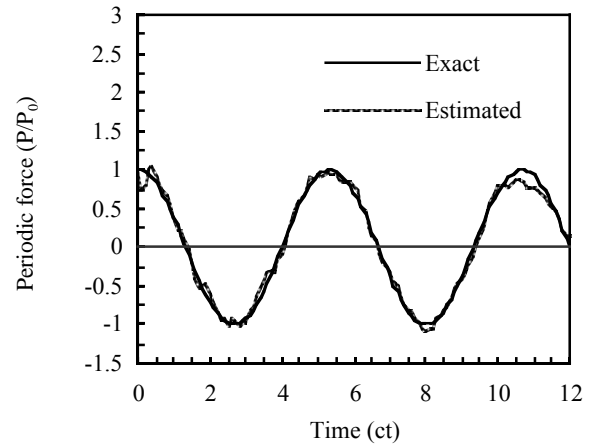

Fig. 3. Periodic load estimation based on velocity measurements of 5 percent noise level at two points $A$ and $C$ ($B$=13.5)

Numerical experiments are concerned with the periodic load. The recognized time-force curves using one-point and two-point velocity measurements of 5% noise level are respectively illustrated in Figs. 2 and 3. It is seen from Fig. 2 that except for an apparent disagreement in the closing time range, the estimation using a single point measurement agrees very well with the true input force history. In contrast, Fig. 3 shows that the prediction based on two-point measurements is accurate in the whole time range.

To provide more insights into the affect of measurements on estimation, Figs. 4, 5 and 6 plot the predicted periodic load using 10% noisy data of velocity or displacement at one point and two points, respectively. Overall, all estimates give good agreement. It is found from Figs. 4 and 5 that the estimations based on a single point measurement always lose big accuracy in the ending time range. In particular, it is observed that the prediction with one-point displacement measurement encounters heavier loss of accuracy in the final time although overall estimation is very smooth. While, the estimation using two-point velocity measurements as shown in Fig. 6 remains accurate in the whole time range. Also the summation of absolute errors by two-point velocity measurements is less than that either by two-point velocity and displacement measurements or by single-point velocity measurement.

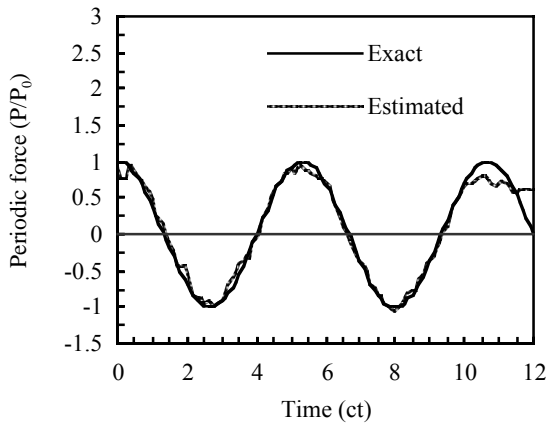

Fig. 4. Periodic load estimation based on velocity measurement of 10 percent noise level at one point $C$ ($B$=6.4)

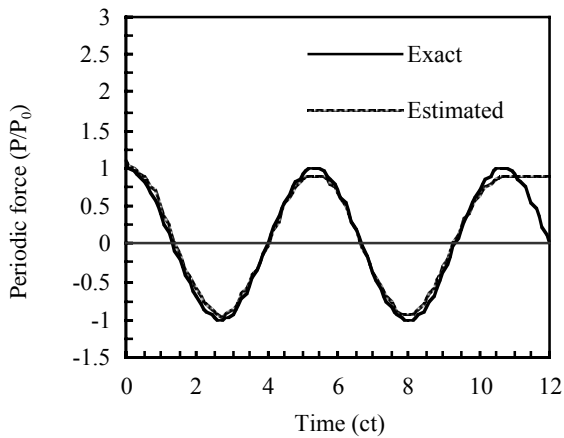

Fig. 5. Periodic load estimation based on displacement measurement of 10 percent noise level at one point $C$ ($B$=7.3)

Furthermore, Fig. 7 shows the recognized periodic force-time curve by using 20% noisy measurements of velocity at point $C$ and of displacement at point $A$. Fig. 8 displays the estimated force curve by using two-point displacement measurements of 20% noisy level. It is found that the accuracy loss is again encountered in these two estimations in the closing range as in the previous cases using one-point measurement, slightly more obvious for two-point displacement case. In contrast, Fig. 9 reveals that the prediction based on two-point velocity measurements is still accurate overall with some amplitude attenuation at the last phase. Therefore, in the case of periodic load, the estimation based on the two-point velocity measurements always works better and is preferred.

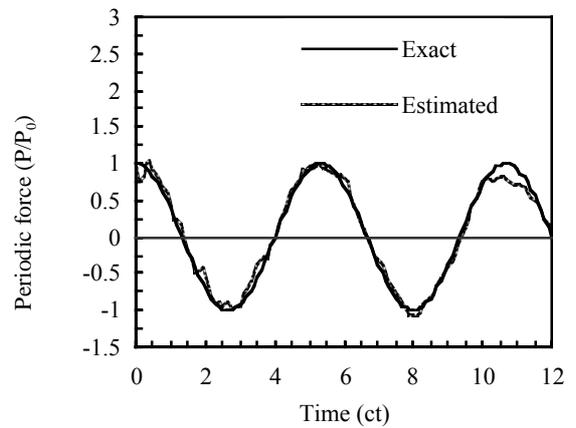

Fig. 6. Periodic load estimation based on velocity measurements of 10 percent noise level at two points $A$ and $C$ ($B$=17.6)

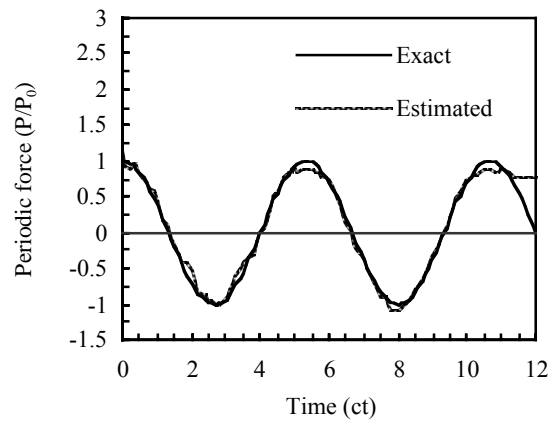

Fig. 7. Periodic load estimation based on displacement and velocity measurements of 20 percent noise level respectively at points $A$ and $C$ ($B$=21.7)

To demonstrate the ability of the present inverse method to the heavy noise measurements, the noise level is raised to 40%. The estimation using two-point velocity measurements is illustrated in Fig. 9. It is observed that the prediction encounters visible amplitude attenuation. This is because the larger noise level degrades the estimation. However, the overall estimate still well reflects the true nature of periodic load.

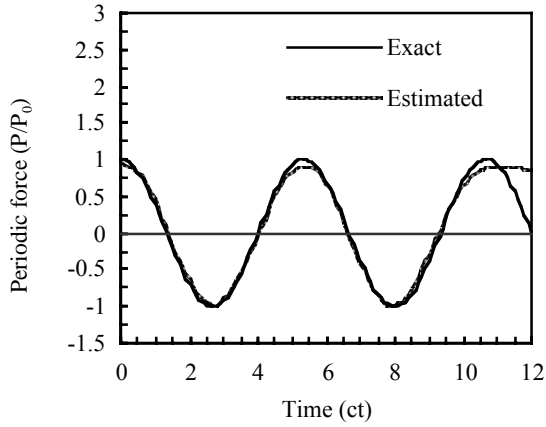

Fig. 8. Periodic load estimation based on displacement measurements of 20 percent noise level at two distinct points *A* and *C* (*B*=19.4)

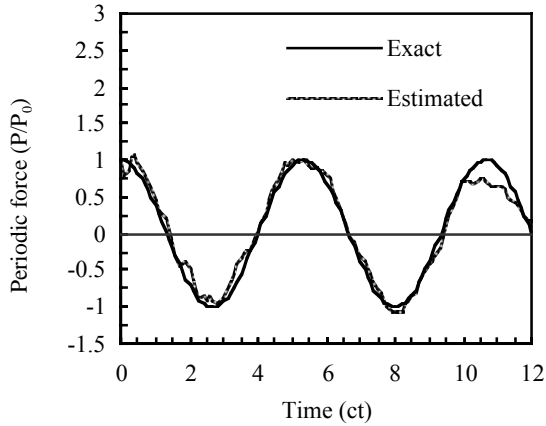

Fig. 9. Periodic load estimation based on velocity measurements of 20 percent noise level at two distinct points *A* and *C* (*B*=23.2)

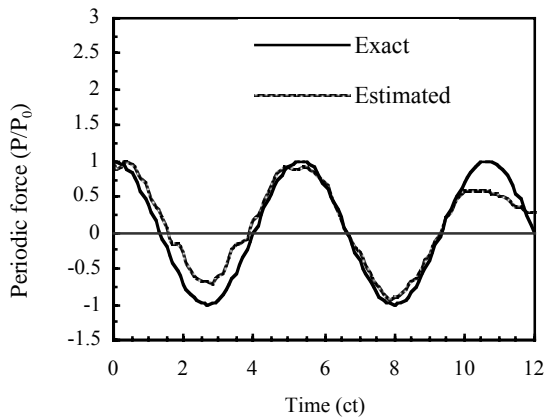

Fig. 10. Periodic load estimation based on velocity measurements of 40 percent noise level at two distinct points *A* and *C* (*B*=105.5)

We also tested the present approach to the estimation of impact load such as the Heaviside impact. The results illustrated in the following Figs. 11 -13 also verified the accuracy, efficiency and reliability of this strategy for impact load where high frequency components play an important role.

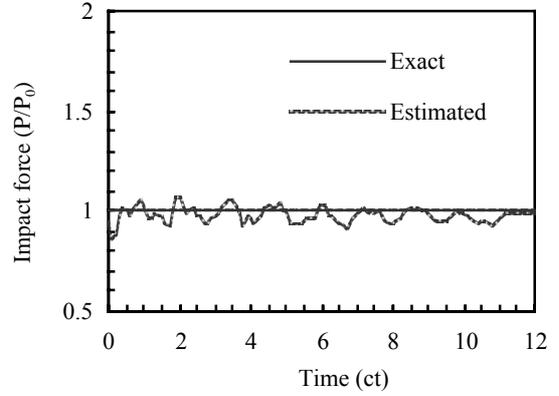

Fig. 11. Heaviside impact force estimation based on velocity measurement of 20 percent noise level at one point *C* (*B*=38.4)

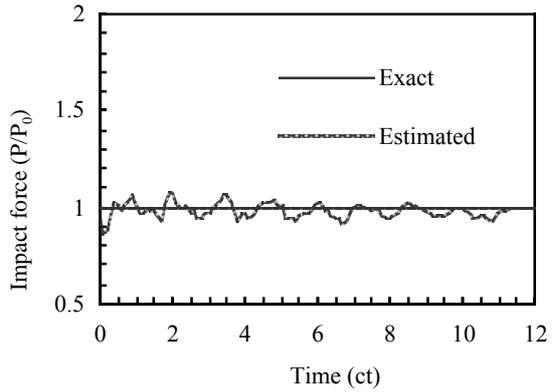

Fig. 12. Heaviside impact force estimation based on displacement measurement of 20 percent noise level at one point *C* (*B*=40.8)

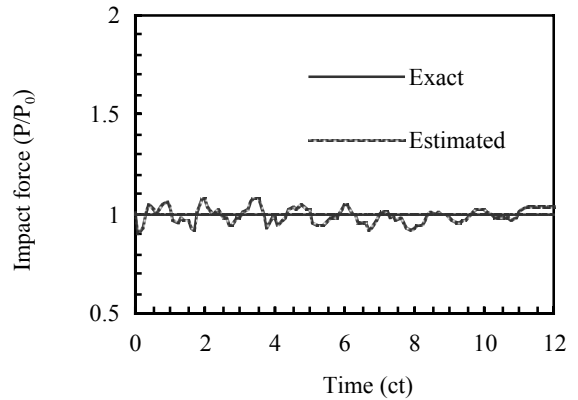

Fig. 13. Heaviside impact force estimation based on both displacement and velocity measurements of 20 percent noise level at two distinct points *A* and *C* (*B*=43.4)

It is found from these experimental results that unlike the harmonic load, the measurement type and number have no evident affect on the estimation accuracy in the Heaviside load case.

## 5. Conclusions

The foregoing numerical experiments demonstrated that the DRBEM in conjunction with the dynamic programming filter is an accurate, robust and computationally efficient methodology to identify input load based on velocity and displacement measurements of different noise level. It was found that the present approach is insensitive to measurement errors and can give good estimation even using heavily noisy data. For the present periodic load, the velocity measurements in two distinct points produce better prediction in general. Also, it is noted that the performances are not very sensitive to regularization parameter.

The present combined approach is mathematically simple and easy to computer programming. The dynamic programming filter is found to satisfy the criteria for a competitive inverse method proposed by Beck et al. [13], while the DRBEM is a powerful technique to transfer a variety of continuous dynamic problems to discrete systems. Therefore, the coupling application of these two methods should be extremely promising for practical inverse elastodynamic analysis.

**Acknowledgements**:
This work was carried out as a part of the research program supported by the Japan Society for the Promotion of Science. Additional financial support was provided as Grant-in-Aid for JSPS fellows by the Ministry of Education, Science, Sports and Culture, Japan.

## References

1. Hensel, E., *Inverse Theory and Applications for Engineers*, Prentice Hall, New Jersey 1991.
2. Kurpisz, K. and Nowak, A.J., *Inverse Thermal Problems*, Comp. Mech. Publ., Southampton, UK, 1995.
3. Trujillo, D.M. and Busby, H.R., *Practical Inverse Analysis in Engineering*, CRC Press, New York, 1997.
4. Trujillo, D.M. and Busby, H.R., Investigation of a technique for the differentiation of empirical data, *ASME J. Dynamic Systems, Meas. Contr.*, **105**, 1983, pp. 200-203.
5. Partridge, P.W., Brebbia, C.A. and Wrobel, L.W., *The Dual Reciprocity Boundary Element Method*, Comput. Mech. Publ., Southampton, UK, 1992.
6. Katsikadelis, J.T. and Nerantzaki, M.S., The boundary element method for nonlinear problems, *Engineering Analysis with Boundary Elements*, **23,** 1999, pp. 365-373.
7. Ingham, D.B., Identification of thermal properties of heat conducting materials, in *Boundary Integral Formulations for Inverse Analysis*, ed. D.B. Ingham and L.C. Wrobel, Comput. Mech. Publ., Southampton, UK, 1997, pp. 1-33.
8. Busby, H.R. and Trujillo, D.M., Solution of an inverse dynamic problems using an eigenvalue reduction technique, *Computers & Structures*, **25**-1 , 1987, pp. 109-117.
9. Zhong, W.X. and Williams, F.W., A precise time step integration method, *J. Mech. Engng. Sci.*, Part C, **208**-6 , 1994, pp. 427-430.
10. Kondou, K., *Theory of Vibration* (in Japanese), Baihu Kan Press, Tokyo, 1993.
11. Wrobel, L.C. and Brebbia, C. A. , The dual reciprocity boundary element formulation for nonlinear diffusion problems, *Computer Methods in Applied Mechanics and Engineering*, **65**, 1987, pp. 147-164.
12. Press, W.H., Teukolsky, S.A., Vetterling, W.T. and Flannery, B.P., *Numerical Recipes in Fortran*, Cambridge University Press, 1992.
13. Beck, J.V., Blackwell, B. and St.Clair, C.R., *Inverse Heat Conduction: Ill-Posed Problems*, Wiley Intersc., New York, 1985.